\documentclass[twocolumn]{revtex4}
\usepackage{amsfonts}
\usepackage{amsmath}
\usepackage{amssymb}
\usepackage{charter}
\usepackage{graphicx}

\input{tcilatex}
\begin{document}

\title{Generating single-photon catalyzed coherent states with
quantum-optical catalysis}
\author{Xue-xiang Xu$^{1,\dag }$ and Hong-chun Yuan$^{2}$}
\affiliation{$^{1}$College of Physics and Communication Electronics,Jiangxi Normal
University, Nanchang 330022, China\\
$^{2}$College of Electrical and Optoelectronic Engineering, Changzhou
Institute of Technology, Changzhou 213002, China\\
$^{\dag }$Corresponding author: xuxuexiang@jxnu.edu.cn}

\begin{abstract}
We generate single-photon catalyzed coherent states (SPCCSs) by means of
quantum-optical catalysis based on the beam splitter (BS) or the parametric
amplifier (PA). These states are obtained in one of the BS (or PA) output
channels if a coherent state and a single-photon Fock state are present in
two input ports and a single photon is registered in the other output port.
The success probabilities of the detection (also the normalization factors)
are discussed, which is different for BS and PA catalysis. In addition, we
prove that the generated states catalyzed by BS and PA devices are actually
the same quantum states after analyzing photon number distribution of the
SPCCSs. The quantum properties of the SPCCSs, such as sub-Poissionian
distribution, anti-bunching effect, quadrature squeezing effect, and the
negativity of the Wigner function are investigated in detail. The results
shows that the SPCCSs are non-Gaussian states with an abundance of
nonclassicality, which can provide the quantum advantages for quantum
technological tasks.

\end{abstract}

\maketitle

\section{Introduction}

According to von Neumann's projection principle \cite{1}, when some
measurement is performed on one subsystem of the quantum-mechanically
correlated system, the effect of the measurement outcome appears in the
other subsystem. In particular, when a correlated two-mode optical field is
prepared in an entangled state of two subsystems and the measurement is
performed on one subsystem, then the quantum state of the other subsystem
can be reduced to a new state \cite{2,3}. The unobserved output state which
depends on the measurement outcome is called as the conditional output
state. The measurement performed to obtain the conditional output state is
named as conditional measurement. Conditional measurement may be a fruitful
method for quantum-state manipulation and engineering. Many nonclassical
states, such as Schrodinger-cat-like state \cite{4}, arbitrary Fock states 
\cite{5}, photon-subtracted traditional quantum states \cite{6}, arbitrary
superposition of coherent states\cite{7}, and arbitrary multimode entangled
states \cite{8} or as well as other nonclassical states\cite{8a}, have been
generated by conditional measurements theoretically or experimentally.

In general, two quantum states in the two output ports of the lossless beam
splitter (BS) and nondegenerate parametric amplifier (PA) are
quantum-mechanically correlated with each other, even if two input states
are not correlated. Hence the BS and the PA are key optical devices to
obtain quantum-mechanically correlated state \cite{9,9a}. If appropriate
measurement, such as homodyne measurement and photon counting, is employed
in one of the output ports, then conditional quantum state is generated in
other output port. Among the schemes of conditional measurement, the most
feasible strategy is \textquotedblleft quantum-optical
catalysis\textquotedblright , proposed by Lovvky and Mlynek \cite{10}. They
generated a coherent superposition state $t\left\vert 0\right\rangle +\alpha
\left\vert 1\right\rangle $ by conditional measurement on a BS. This state
was generated in one of the BS output channels if a coherent state $%
\left\vert \alpha \right\rangle $ and a single-photon Fock state $\left\vert
1\right\rangle $ are present in two input ports and a single photon is
registered in the other BS output. They called this transformation
\textquotedblleft quantum-optical catalysis\textquotedblright\ because the
single photon itself remains unaffected but facilitate the conversion of the
target ensemble. Subsequently, Bartley \textit{et al.} \cite{11} used
\textquotedblleft quantum-optical catalysis\textquotedblright\ to generate
multiphoton nonclassical state, exhibiting a wide range of nonclassical
phenomena. Recently, we operated \textquotedblleft quantum-optical
catalysis\textquotedblright\ on each mode of the two-mode squeezed vacuum
state and generated a non-Gaussian two-mode quantum state with higher
entanglement \cite{12}.

In 1997, Ban had derived the equivalence between BS and PA in conditional
quantum measurement \cite{13}. In other words, the conditional output of the
BS is equal to that of the PA under some conditions. This inspire us to
reconsider the works of Lovvky and Bartley \cite{10}. In this work, we
theoretically generate a kind of single-photon catalyzed coherent states
(SPCCSs) by operating single-photon quantum-optical catalysis on a coherent
state based on two kinds of quantum optical devices (BS and PA). In
addition, we investigate some nonclassical properties of the generated
states.

The paper is organized as follows. In Sec. II, using the BS and the PA as
the basic devices we generate single-photon catalyzed coherent states
(SPCCSs). The theoretical schemes are proposed and their detection
probabilities are discussed. In Sec. III, we prove that the generated states
by BS and PA devices are the same quantum states if the catalysis parameters
are chosen appropriately. In Sec. IV, the nonclassical properties of the
generated states, such as sub-Poissionian distribution, anti-bunching
effect, and quadrature squeezing effect are investigated. Subsequently, the
negativity of the Wigner functions of the SPCCSs is investigated in Sec. V.
The main results are summarized in Sec.VI.

\section{Single-photon catalyzed coherent state}

Of all states of the radiation field, the coherent states are the most
important and arise frequently in quantum optics \cite{14}. Coherent state
are generally accepted to be the most classical of the quantum states \cite%
{15}. In this section, we use a coherent state (CS) as the initial state and
make \textquotedblleft quantum-optical catalysis\textquotedblright\ to
induce some nonclassical states. These states will exhibit an abundance of
nonclassical properties distinguished from that of the coherent states, as
shown below.

As shown schematically in Fig.1, based on two kinds of optical devices (BS
or PA) \cite{16,17}, we generate the SPCCSs $\left\vert \alpha
_{c}\right\rangle $ ($\left\vert \alpha _{c}\right\rangle _{BS}$ and $%
\left\vert \alpha _{c}\right\rangle _{PA}$). If an coherent state $%
\left\vert \alpha \right\rangle $ and a single-photon Fock state $\left\vert
1\right\rangle $ are present in the two input ports of one optical device
and a single photon $\left\vert 1\right\rangle $\ is registered in one
output port, then a catalyzed state $\left\vert \alpha _{c}\right\rangle $
can be generated in the other output channel. Since the quantum-optical
catalysis is a process of postselection, the success probabilities of
detection are analyzed numerically. 
\begin{figure}[tbp]
\label{Fig1} \centering\includegraphics[width=1.0\columnwidth]{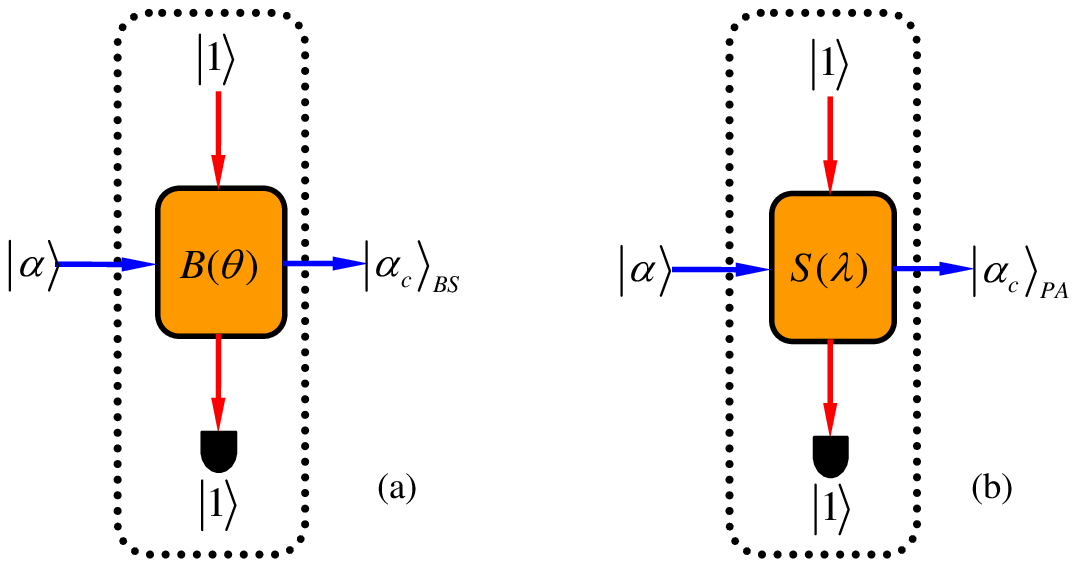}
\caption{(color online) Condtional generation of the SPCCSs $\left\vert 
\protect\alpha _{c}\right\rangle $ with quantum-optical catalysis from a
coherent state $\left\vert \protect\alpha \right\rangle $, where the
interaction parameter is the catalysis parameter $\Lambda $. (a) Using the
BS device described with a operator $B\left( \protect\theta \right) $ and $%
\Lambda =r^{2}=\sin ^{2}\protect\theta $; (b) Using the PA device describe
with a operator $S\left( \protect\lambda \right) $ and $\Lambda =\protect%
\kappa ^{2}=\tanh ^{2}\protect\lambda $. Under the constraint $\Lambda
=r^{2}=\protect\kappa ^{2}$ is satisfied, the conditional output state of
the BS ($\left\vert \protect\alpha _{c}\right\rangle _{BS}$) is equal to
that of the PA ($\left\vert \protect\alpha _{c}\right\rangle _{PA}$).}
\end{figure}

\subsection{SPCCS prepared by the BS device}

In Fig.1(a), the role played by the lossless beam splitter (BS) upon the
input state $\left\vert \psi _{in}\right\rangle =\left\vert \alpha
\right\rangle _{a}\left\vert 1\right\rangle _{b}$ results in the output
state $\left\vert \psi _{out}\right\rangle =B\left( \theta \right)
\left\vert \alpha \right\rangle _{a}\left\vert 1\right\rangle _{b}$, where $%
B\left( \theta \right) =\exp \left[ \theta \left( a^{\dag }b-ab^{\dag
}\right) \right] $ corresponds to the unitary operator of the adjustable BS $%
B\left( \theta \right) \ $in terms of the creation (annihilation) operator $%
a^{\dag }(a)$ and $b^{\dag }(b)$ for modes $a$ and $b$, which fulfill $%
Ba^{\dag }B^{\dag }=a^{\dag }t-b^{\dag }r\ $and $Bb^{\dag }B^{\dag }=a^{\dag
}r+b^{\dag }t$ with $r=\sin \theta $ and\ $t=\cos \theta $ ($\theta \in %
\left[ 0,\pi /2\right] $).\ After registering single-photon in the output
mode $b$, a SPCCS is obtained in $a$ channel%
\begin{eqnarray}
\left\vert \alpha _{c}\right\rangle _{BS} &=&\frac{1}{\sqrt{p_{BS}}}\left.
_{b}\left\langle 1\right\vert B\left( \theta \right) \left\vert \alpha
\right\rangle _{a}\left\vert 1\right\rangle _{b}\right.  \notag \\
&=&c_{0}\allowbreak \left\vert t\alpha \right\rangle +c_{1}a^{\dag
}\left\vert t\alpha \right\rangle ,  \label{1-1}
\end{eqnarray}%
where $\left\vert t\alpha \right\rangle $\ is a new coherent state and $%
c_{0}=te^{-r^{2}\left\vert \alpha \right\vert ^{2}/2}/\sqrt{p_{BS}}%
,c_{1}=-r^{2}\alpha e^{-r^{2}\left\vert \alpha \right\vert ^{2}/2}/\sqrt{%
p_{BS}}$. In addition, the normalization factor $p_{BS}=e^{-r^{2}\left\vert
\alpha \right\vert ^{2}}I_{0}\left( \alpha ,r^{2}\right) $ is the success
probability heralded by the detection of a single photon at the mode, where
the function $I_{0}\left( \alpha ,r^{2}\right) $ is defined in appendix.
Obviously, $I_{0}\left( \alpha ,0\right) =1$ and $I_{0}\left( \alpha
,1\right) =\allowbreak \left\vert \alpha \right\vert ^{2}$. The success
probability $p_{BS}$ of obtaining the SPCCSs $\left\vert \alpha
_{c}\right\rangle _{BS}$ is plotted in the ($\left\vert \alpha \right\vert
,r^{2}$) parameter space in Fig.2 (a) and (c).

\subsection{SPCCS prepared by the PA device}

In Fig.1(b), the role played by the nondegenerate parametric amplifier
(NOPA) upon the input state $\left\vert \psi _{in}\right\rangle =\left\vert
\alpha \right\rangle _{a}\left\vert 1\right\rangle _{b}$ results in the
output state $\left\vert \psi _{out}\right\rangle =S\left( \lambda \right)
\left\vert \alpha \right\rangle _{a}\left\vert 1\right\rangle _{b}$, where $%
S\left( \lambda \right) =\exp \left[ \lambda \left( a^{\dag }b^{\dag
}-ab\right) \right] $ corresponds to the two-mode squeezed operator with
real squeezing paramter $\lambda $, which fulfill $Sa^{\dag }S^{\dag
}=\varkappa ^{-1}a^{\dag }-\kappa \varkappa ^{-1}b,$ \ $Sb^{\dag }S^{\dag
}=\varkappa ^{-1}b^{\dag }-\kappa \varkappa ^{-1}a$ with $\kappa =\tanh
\lambda $ and\ $\varkappa =\cosh ^{-1}\lambda $ ($\lambda \in \left[
0,\infty \right) $). Similarly, when single-phton is detected in the $b$
channel, a SPCCS is obtained in the $a$ channel 
\begin{eqnarray}
\left\vert \alpha _{c}\right\rangle _{PA} &=&\frac{1}{\sqrt{p_{PA}}}\left.
_{b}\left\langle 1\right\vert S\left( \lambda \right) \left\vert \alpha
\right\rangle _{a}\left\vert 1\right\rangle _{b}\right.  \notag \\
&=&d_{0}\left\vert \varkappa \alpha \right\rangle +d_{1}a^{\dag }\left\vert
\varkappa \alpha \right\rangle ,  \label{1-2}
\end{eqnarray}%
where $\left\vert \varkappa \alpha \right\rangle $\ is also a coherent state
and $d_{0}=\varkappa ^{2}e^{-\kappa ^{2}\left\vert \alpha \right\vert
^{2}/2}/\sqrt{p_{PA}}$ and $d_{1}=-\varkappa \kappa ^{2}\alpha e^{-\kappa
^{2}\left\vert \alpha \right\vert ^{2}/2}/\sqrt{p_{PA}}$. In addition, the
normalization factor $p_{PA}=\left( 1-\kappa ^{2}\right) e^{-\kappa
^{2}\left\vert \alpha \right\vert ^{2}}I_{0}\left( \alpha ,\kappa
^{2}\right) $ is the success probability of such event. The success
probability $p_{PA}$ of obtaining the SPCCSs $\left\vert \alpha
_{c}\right\rangle _{PA}$ are plotted in the ($\left\vert \alpha \right\vert
,\kappa ^{2}$) parameter space in Fig.2 (b) and (d). 
\begin{figure}[tbp]
\label{Fig6-1} \centering\includegraphics[width=1.0\columnwidth]{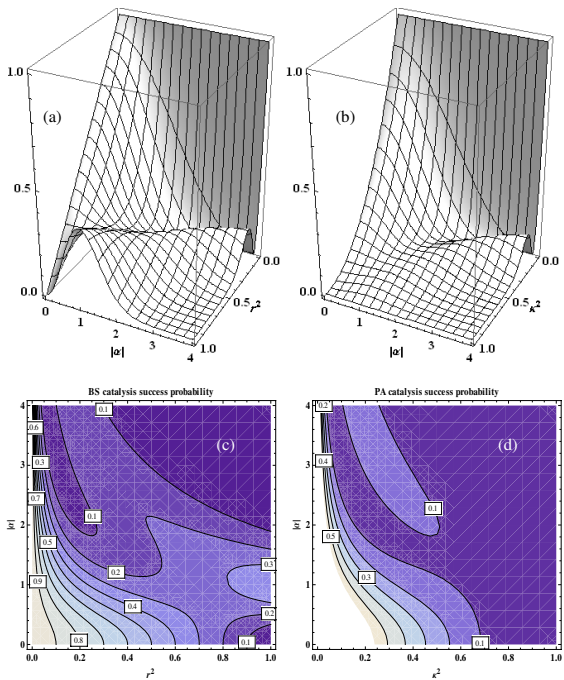}
\caption{(Color online) The success probabilities of detection (a) and (c)
for $p_{BS}$ as a function of interaction parameter $\left\vert \protect%
\alpha \right\vert $ and $r^{2}$; (b) and (d) for $p_{PA}$ as a function of
interaction parameter $\left\vert \protect\alpha \right\vert $ and $\protect%
\kappa ^{2}$.}
\end{figure}

\section{Equivalent effect of BS and PA in preparing the SPCCS}

From Eqs.(\ref{1-1}) and (\ref{1-2}), we find that the SPCCSs ($\left\vert
\alpha _{c}\right\rangle _{BS}$ and $\left\vert \alpha _{c}\right\rangle
_{PA}$) are the coherent superposition state of a coherent state (Gaussian, $%
\left\vert t\alpha \right\rangle $ or $\left\vert \varkappa \alpha
\right\rangle $) and a single-photon-added coherent state (non-Gaussian, $%
a^{\dag }\left\vert t\alpha \right\rangle $ or $a^{\dag }\left\vert
\varkappa \alpha \right\rangle $) with certain ration. Observing the forms
of the SPCCSs ($\left\vert \alpha _{c}\right\rangle _{BS}$ and $\left\vert
\alpha _{c}\right\rangle _{PA}$) in Eqs.(\ref{1-1}) and (\ref{1-2}), one can
find that they are all the superposition states of a coherent state and a
photon-added coherent state. A question on whether there exist a kind of
link between $\left\vert \alpha _{c}\right\rangle _{BS}$ and $\left\vert
\alpha _{c}\right\rangle _{PA}$ is naturally arisen? Here we will discuss
this question.

Expanding $\left\vert \alpha _{c}\right\rangle _{BS}$ and $\left\vert \alpha
_{c}\right\rangle _{PA}$ into the Fock basis, we have the same form as
follows%
\begin{equation}
\left\vert \alpha _{c}\right\rangle _{\Lambda }=\sum_{n=0}^{\infty }\omega
_{n}^{\Lambda }\left\vert n\right\rangle ,  \label{2-1}
\end{equation}%
with the coefficients%
\begin{eqnarray}
\omega _{n}^{\Lambda } &=&\frac{\alpha ^{n}e^{-\left( 1-\Lambda \right)
\left\vert \alpha \right\vert ^{2}/2}}{\sqrt{n!I_{0}\left( \alpha ,\Lambda
\right) }}  \notag \\
&&\times \sqrt{1-\Lambda }^{n-1}\left( \allowbreak 1-\Lambda -n\Lambda
\right) ,  \label{2-2}
\end{eqnarray}%
leading to the photon number distribution (PND) of the catalyzed states $%
P_{\Lambda }\left( n\right) =\left\vert \omega _{n}^{\Lambda }\right\vert
^{2}$. It is found that $\left\vert \alpha _{c}\right\rangle _{BS}$ and $%
\left\vert \alpha _{c}\right\rangle _{PA}$ are the same quantum states as
long as the condition $\Lambda =r^{2}=\kappa ^{2}$ is satisfied. Here is
equate to $r^{2} $ for the BS case and $\kappa ^{2}$ for the PA case. By
analyzing the construction of the generated states, we verify that the SPCCS
generated on the BS is equal to that generated on the PA if we choose the
appropriate catalysis parameters.

It is obvious to see that the SPCCSs are characterized by two parameters,
i.e. the input parameter $\alpha $ and the catalysis parameter $\Lambda $.
By adjusting the parameters, the coefficients may be modulated, generating a
wide range of nonclassical phenomena, as seen in the Sec.IV. Especially,
when $\Lambda =0$, the SPCCS reduces to the input CS $\left\vert \alpha
\right\rangle $ with $P_{0}\left( n\right) =e^{-\left\vert \alpha
\right\vert ^{2}}\allowbreak \left\vert \alpha \right\vert ^{2n}/n!$
(Poissonian distribution) \cite{18}; while $\Lambda =1$, the SPCCS reduces
to $\left\vert 1\right\rangle $ with $P_{1}\left( n\right) =\delta _{1,n}$.
In Fig.3 we plot the PNDs of the coherent state $\left\vert \alpha
\right\rangle $ with $\left\vert \alpha \right\vert =1$ and the SPCCS with $%
\left\vert \alpha \right\vert =1$ and $\Lambda =0.7$. 
\begin{figure}[tbp]
\label{Fig7-1} \centering\includegraphics[width=0.8\columnwidth]{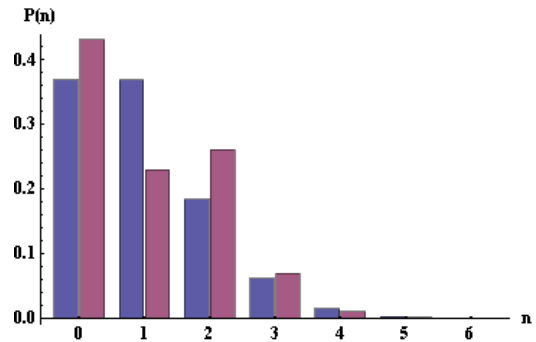}
\caption{(Color online) Photon-number distributions of (a) Coherent state $%
\left\vert \protect\alpha \right\rangle $ with $\left\vert \protect\alpha %
\right\vert =1$ (blue bars) and (b) the SPCCS with $\left\vert \protect%
\alpha \right\vert =1$ and $\Lambda =r^{2}=\protect\kappa ^{2}=0.7$ (purple
bars).}
\end{figure}

\section{ Nonclassical properties of the SPCCS}

Quantum states of light can be classified according to their statistical
properties. They are usually compared to a reference state, namely, the
coherent state \cite{19}. Hence in this section, we compare the SPCCSs with
the origin coherent state and discuss the nonclassical properties of the
SPCCSs.

\subsection{Sub-Poissionian distribution and anti-bunching effect}

In this subsection, we examine the Mandel Q factor \cite{20} 
\begin{equation}
Q=\frac{\left\langle a^{\dagger 2}a^{2}\right\rangle }{\left\langle
a^{\dagger }a\right\rangle }-\left\langle a^{\dagger }a\right\rangle
\label{3-1}
\end{equation}%
and measure the second-order autocorrelation function \cite{21} 
\begin{equation}
g^{\left( 2\right) }\left( 0\right) =\frac{\left\langle a^{\dagger
2}a^{2}\right\rangle }{\left\langle a^{\dagger }a\right\rangle ^{2}}.
\label{3-2}
\end{equation}%
The distribution is Poissonian when $Q=0$, and super- (sub-) Poissonian if $%
Q>0$ ($Q<0$), while the effect is antibunching when $g^{\left( 2\right)
}\left( 0\right) <1$\ (strictly nonclassical), and bunching (superbunching)
if $1\leqslant g^{\left( 2\right) }\left( 0\right) \leqslant 2$ ($g^{\left(
2\right) }\left( 0\right) >2$). For a coherent state, $g^{\left( 2\right)
}\left( 0\right) =1$ corresponds to $Q=0$ (Poissonian statistics).

The variations of $Q$ are depicted in Fig.4. Mandel Q factor as a function
of for $\left\vert \alpha \right\vert =1,2,3$\ is plotted in Fig.4(a). The
feasibility regions of super-Poisson and sub-Poisson distribution are shown
in Fig.4(b). For $\Lambda \rightarrow 0$, $Q$ \ approaches the values for a
coherent state, i.e. $Q=0$; while $\Lambda \rightarrow 1$, $Q$ \ approaches
the values for a single-photon Fock state, i.e. $Q=-1$. Similarly, the
second-order autocorrelation function as a function of for $\left\vert
\alpha \right\vert =1,2,3$\ is plotted in Fig.5(a) and the feasibility
regions of antibunching, bunching and superbunching are shown in Fig.5(b).
For $\Lambda \rightarrow 0$, $g^{\left( 2\right) }\left( 0\right) $ \
approaches the values for a coherent state, i.e. $g^{\left( 2\right) }\left(
0\right) =1$; while $\Lambda \rightarrow 1$, $g^{\left( 2\right) }\left(
0\right) $ \ approaches the values for a single-photon Fock state, i.e. $%
g^{\left( 2\right) }\left( 0\right) =0$. It is found that there may present
sub-Poissonian and antibunching effect in a wide range of interaction
parameters for the SPCCSs. In the limiting case, when $\Lambda =0$, the
states corresponding to CS $\left\vert \alpha \right\rangle $, then $Q=0$
and $g^{\left( 2\right) }\left( 0\right) =1;$while for $\Lambda =1$, the
states corresponding to single-photon Fock state $\left\vert 1\right\rangle $%
, then $Q=-1$ and $g^{\left( 2\right) }\left( 0\right) =0$. 
\begin{figure}[tbp]
\label{Fig5} \centering\includegraphics[width=0.8\columnwidth]{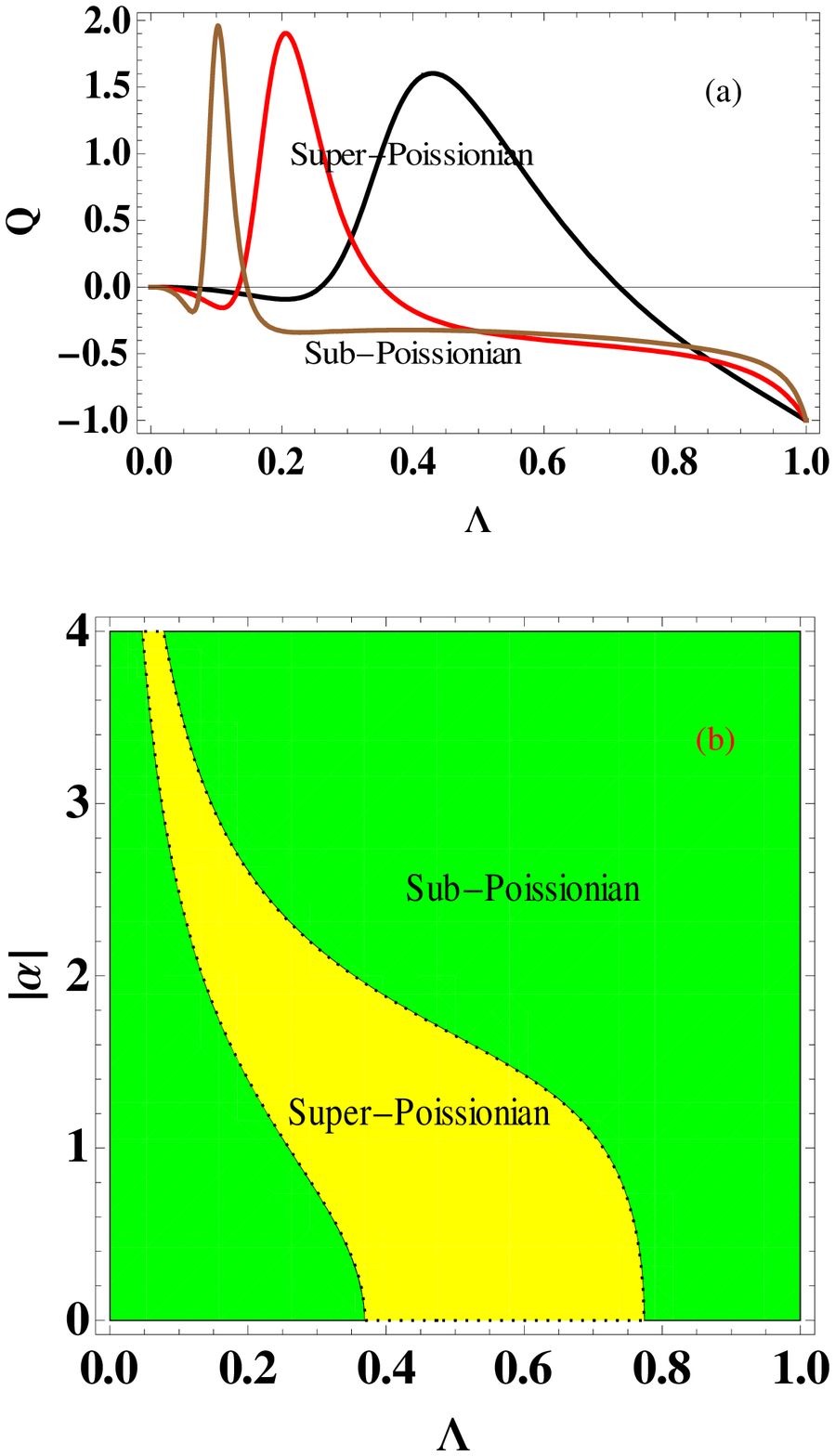}
\caption{(Color online) (a) Mandel Q parameter\ as a function of the
catalysis parameter $\Lambda $ for different $\left\vert \protect\alpha %
\right\vert $, where the black, red, and brown lines are corresponding to $%
\left\vert \protect\alpha \right\vert =1$, $\left\vert \protect\alpha %
\right\vert =2$, and $\left\vert \protect\alpha \right\vert =3$,
respectively. (b) The feasibility regions\ in ($\Lambda $, $\left\vert 
\protect\alpha \right\vert $) space showing sub-Poissionian distribution $%
Q<0 $\ (strictly nonclassical), and -Poissionian distribution $Q>0$.}
\end{figure}
\begin{figure}[tbp]
\label{Fig8} \centering\includegraphics[width=0.8\columnwidth]{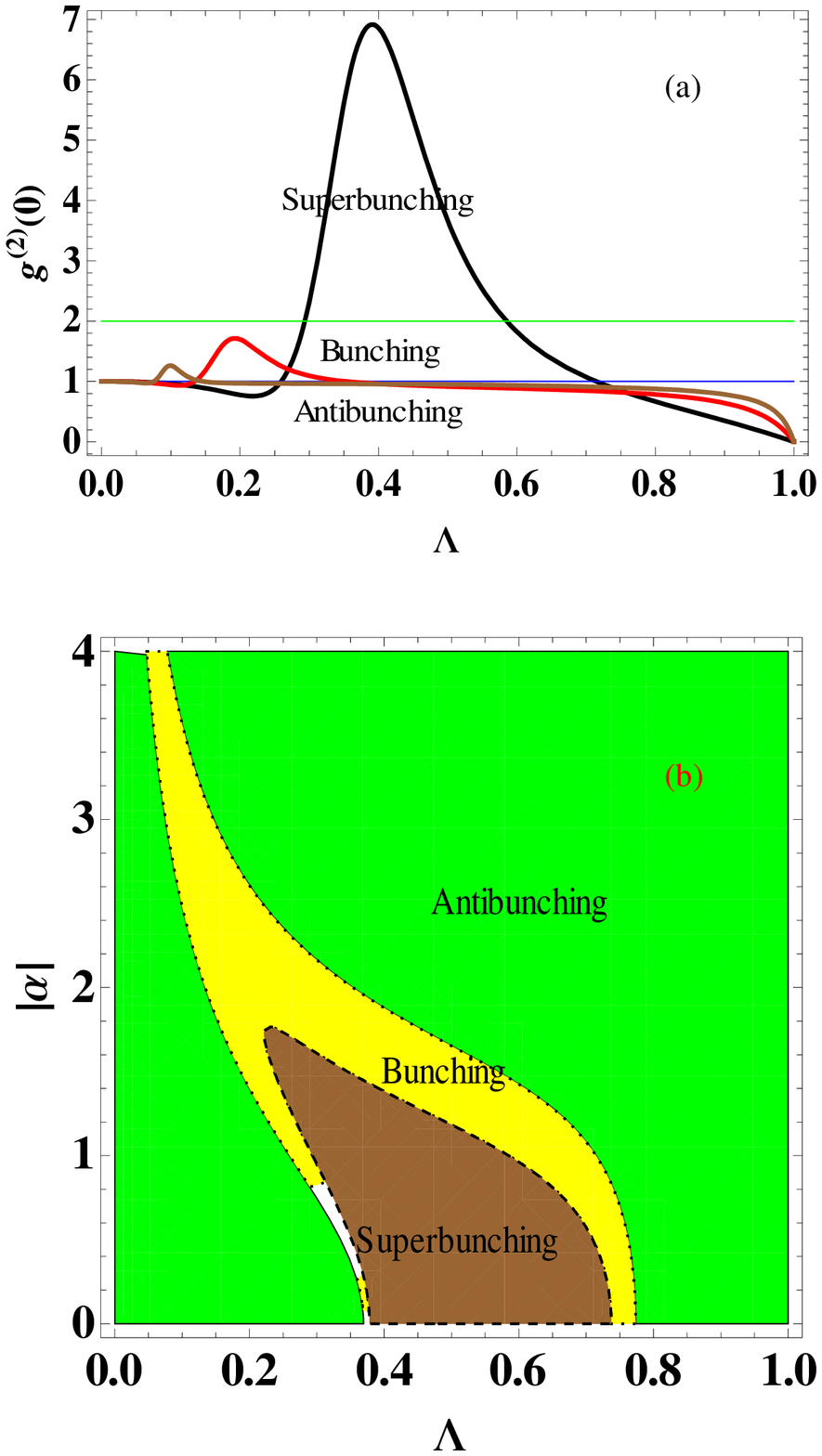}
\caption{(Color online) (a) Second-order autocorrelation function $g^{\left(
2\right) }\left( 0\right) $\ as a function of the catalysis parameter $%
\Lambda $ for different $\left\vert \protect\alpha \right\vert $, where the
black, red, and brown lines are corresponding to $\left\vert \protect\alpha %
\right\vert =1$, $\left\vert \protect\alpha \right\vert =2$, and $\left\vert 
\protect\alpha \right\vert =3$, respectively. (b) The feasibility regions\
in ($\Lambda $, $\left\vert \protect\alpha \right\vert $) space showing
antibunching $g^{\left( 2\right) }\left( 0\right) <1$\ (strictly
nonclassical), bunching $1\leqslant g^{\left( 2\right) }\left( 0\right)
\leqslant 2$, and superbunching $g^{\left( 2\right) }\left( 0\right) >2$.}
\end{figure}

\subsection{Quadrature squeezing effect}

Next, we explore another nonclassical effect, i.e., squeezing of quadrature
amplitude, which is defined from two quadrature operators $X=\left(
a+a^{\dag }\right) /\sqrt{2}$\ and $P=\left( a-a^{\dag }\right) /\left( 
\sqrt{2}i\right) $. Both quadrature variances can expressed as 
\begin{eqnarray}
\left\langle \Delta X^{2}\right\rangle &=&\left\langle a^{\dag
}a\right\rangle -\left\langle a^{\dag }\right\rangle \left\langle
a\right\rangle +\frac{1}{2}  \notag \\
&&+\frac{\left\langle a^{\dag 2}\right\rangle -\left\langle a^{\dag
}\right\rangle ^{2}}{2}+\frac{\left\langle a^{2}\right\rangle -\left\langle
a\right\rangle ^{2}}{2},  \label{3-3}
\end{eqnarray}%
and 
\begin{eqnarray}
\left\langle \Delta P^{2}\right\rangle &=&\left\langle a^{\dag
}a\right\rangle -\left\langle a^{\dag }\right\rangle \left\langle
a\right\rangle +\frac{1}{2}  \notag \\
&&-\frac{\left\langle a^{\dag 2}\right\rangle -\left\langle a^{\dag
}\right\rangle ^{2}}{2}-\frac{\left\langle a^{2}\right\rangle -\left\langle
a\right\rangle ^{2}}{2},  \label{3-4}
\end{eqnarray}%
respectively, as can be seen from their definitions \cite{22}. The
uncertainty relation obeys $\Delta X^{2}\Delta P^{2}\geq 1/4$. For a
coherent (vacuum) state, the variances of $X$\ and $P$\ are equal to $1/2$.
If one of $\Delta X^{2}$ and $\Delta P^{2}$ is smaller than $1/2$, then this
state is squeezing. Moreover, one can also adopt quantum squeezing
quantified in a dB scale through $dB[X]=10\log _{10}\left( \Delta
X^{2}/\Delta X^{2}|_{\left\vert 0\right\rangle }\right) ,dB[P]=10\log
_{10}\left( \Delta P^{2}/\Delta P^{2}|_{\left\vert 0\right\rangle }\right) $%
, i.e., these quadrature variances, relative to their vacuum values of $%
\Delta X^{2}|_{\left\vert 0\right\rangle }=\Delta P^{2}|_{\left\vert
0\right\rangle }=1/2$. Hence, if one of $dB[X]$ and $dB[P]$ is less than $0$%
, this quantum is squeezed state.

In Fig.6(a) we plot the variation of $dB[X]$ as a function of the catalysis
parameter $\Lambda $ for different $\left\vert \alpha \right\vert $. It is
clearly seen that for a given $\left\vert \alpha \right\vert $ there exists
squeezing in $X$ quadrature component in some range of catalysis parameter $%
\Lambda $. The maximum and minimum variances can be found by using the
scientific computing software MATHEMATICA. For the case of $\left\vert
\alpha \right\vert =1$ (see the black line in Fig.7 (a)), the largest
squeezing (corresponding the minimum variance $\Delta X^{2}=3/8$) is
attainable at $\Lambda =0.322185$, below the vacuum noise level of 1/2 by $%
1.25$dB. The maximum variance of 3/2 corresponds to 4.77dB antisqueezing at $%
\Lambda =1$. For the case of $\left\vert \alpha \right\vert =2$ (see the red
line in Fig.7 (b)), the largest squeezing (corresponding the minimum
variance $\Delta X^{2}=3/8$) is attainable at $\Lambda =0.129649$, below the
vacuum noise level of 1/2 by $1.25$dB. The maximum variance of $3/2$
corresponds to 4.77dB antisqueezing at $\Lambda =0.25$ and $\Lambda =1$. For
the case of $\left\vert \alpha \right\vert =3$ (see the brown line in Fig.7
(c)), the largest squeezing (corresponding the minimum variance $\Delta
X^{2}=3/8$) is attainable at $\Lambda =0.0695085$, below the vacuum noise
level of 1/2 by $1.25$dB. The maximum variance of $3/2$ corresponds to
4.77dB antisqueezing at $\Lambda =0.111111$ and $\Lambda =1$. In Fig.6(b),
we plot the contour of $dB[X]$ in ($\Lambda $, $\left\vert \alpha
\right\vert $) plain parameter space. The regions with $dB[X]<0$ show the
squeezing effect. In the limiting case, when $\Lambda =0$, the states
corresponding to CS $\left\vert \alpha \right\rangle $, then $dB[X]=0dB;$%
while for $\Lambda =1$, the states corresponding to single-photon Fock state 
$\left\vert 1\right\rangle $, then $dB[X]=4.77dB$. 
\begin{figure}[tbp]
\label{Fig6} \centering\includegraphics[width=0.8\columnwidth]{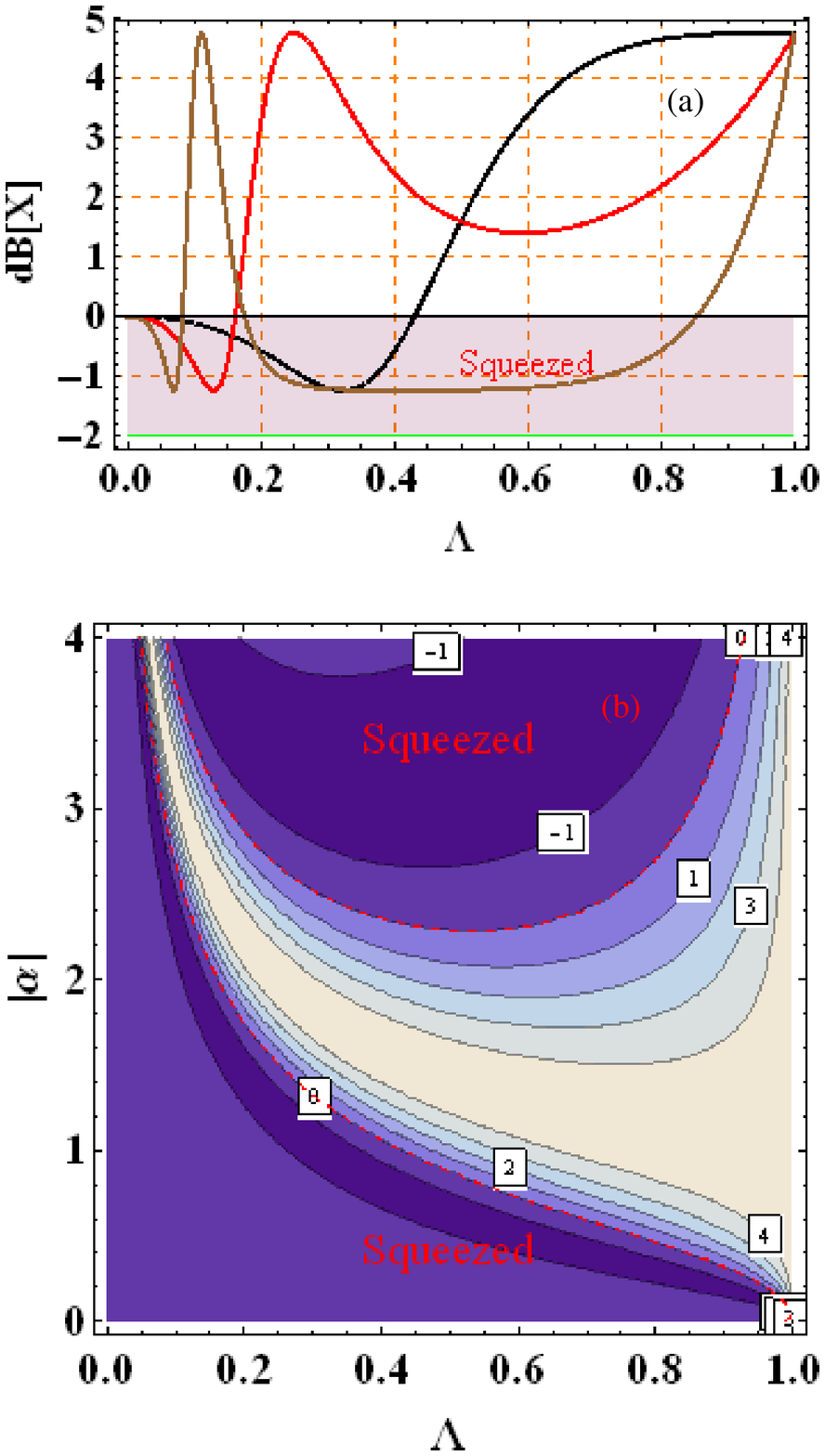}
\caption{(Color online) (a) Quadrature variance of the $X$ component,
relative to the vacuum (unsqueezed) state in units of dB, as a function of
the catalysis parameter $\Lambda $ for different $\left\vert \protect\alpha %
\right\vert $, where the black, red, and brown lines are corresponding to $%
\left\vert \protect\alpha \right\vert =1$, $\left\vert \protect\alpha %
\right\vert =2$, and $\left\vert \protect\alpha \right\vert =3$,
respectively. The grey region shows squeezed. (b) The contourplot of $dB[X]$
in the $\left( \Lambda ,\left\vert \protect\alpha \right\vert \right) $
parameter\ plain space, where squeezing regions correspond the condition of $%
dB[X]<0$.}
\end{figure}

\section{Wigner function of the SPCCS}

The negative Wigner function is a witness of the nonclassicality of a
quantum state \cite{23}. In this section, we derive the analytical
expression of the WF and make numerical analysis for the character of the
SPCCSs. For a single-mode density operator $\rho $, the WF in the coherent
state representation $\left\vert z\right\rangle $ can be expressed as $%
W(\beta )=\frac{2e^{2\left\vert \beta \right\vert ^{2}}}{\pi }\int \frac{%
d^{2}z}{\pi }\left\langle -z\right\vert \rho \left\vert z\right\rangle
e^{-2\left( z\beta ^{\ast }-z^{\ast }\beta \right) }$, where $\beta =\left(
q+ip\right) /\sqrt{2}$\cite{24}. For the SPCCS, the WF is 
\begin{equation}
W(\beta ;\alpha ,\Lambda )=\frac{2F(\beta ;\alpha ,\Lambda )}{\pi
I_{0}\left( \alpha ,\Lambda \right) }e^{-2\left\vert \beta -\sqrt{1-\Lambda }%
\alpha \right\vert ^{2}},  \label{4-1}
\end{equation}%
where the defined function is%
\begin{eqnarray}
&&F(\beta ;\alpha ,\Lambda )  \notag \\
&=&\left( 1-\Lambda \right) -\Lambda \left( 3\Lambda -2\right) \left\vert
\alpha \right\vert ^{2}  \notag \\
&&+\Lambda ^{2}\left( 1-\Lambda \right) \left\vert \alpha \right\vert
^{4}+4\Lambda ^{2}\left\vert \alpha \right\vert ^{2}\left\vert \beta
\right\vert ^{2}  \notag \\
&&-2\Lambda \sqrt{1-\Lambda }\left( 1+\Lambda \left\vert \alpha \right\vert
^{2}\right) \left( \alpha \beta ^{\ast }+\beta \alpha ^{\ast }\right)
\allowbreak .  \label{4-2}
\end{eqnarray}%
Obviously, the WF $W(\beta ;\alpha ,\Lambda )$ in Eq.(\ref{4-1}) is
non-Gaussian in the phase space due to the presence of the term $F(\beta
;\alpha ,\Lambda )/I_{0}\left( \alpha ,\Lambda \right) $. In addition, it
indicates that if $F(\beta ;\alpha ,\Lambda )<0$ then there is negative
region in phase space. In particular, when $\Lambda =0$, Eq.(\ref{4-1}) just
reduces to the WF of the CS $\left\vert \alpha \right\rangle $, i.e., $%
W(\beta ;\alpha ,0)=\frac{2}{\pi }e^{-2\left\vert \beta -\alpha \right\vert
^{2}}$; when $\Lambda =1$, Eq.(\ref{3-2}) just reduces to the WF of
single-photon Fock state $\left\vert 1\right\rangle $, i.e., $W(\beta
;\alpha ,1)=\frac{2}{\pi \allowbreak }(4\left\vert \beta \right\vert
^{2}-1)e^{-2\left\vert \beta \right\vert ^{2}}$. The Wigner distributions
for several SPCCSs with $\alpha =1+i$, $\alpha =2$, and $\alpha =2.7$ are
depicted in phase space in Fig.7. There exist some obvious negative regions
in the phase space, which is an important figure of merit for a non-Gaussian
quantum state. 
\begin{figure}[tbp]
\label{Fig9} \centering\includegraphics[width=1.0\columnwidth]{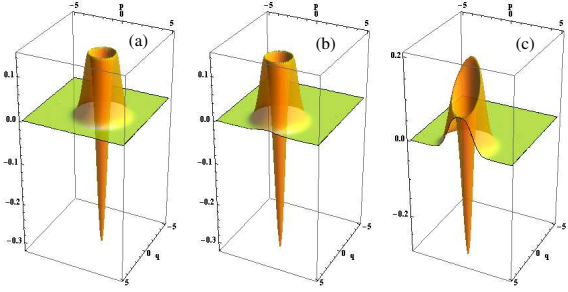}
\caption{(Color online) Wigner functions of the SPCCSs for some different
parameters with (a) $\protect\alpha =1+i$, $\Lambda =0.5$; (b) $\protect%
\alpha =2$, $\Lambda =0.25$; and (c) $\protect\alpha =2.7$, $\Lambda =0.125$%
. Outstanding characteristic is the negativity of the Wigner functions.}
\end{figure}
\begin{figure}[tbp]
\label{Fig9x} \centering\includegraphics[width=0.9\columnwidth]{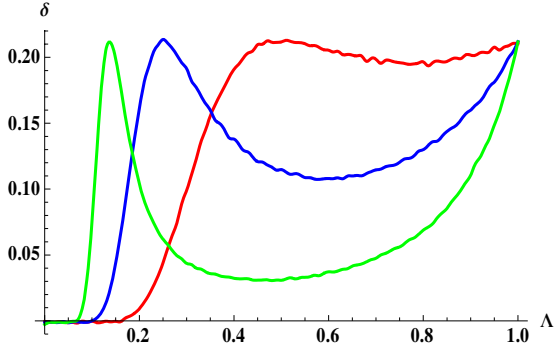}
\caption{(Color online) Negative volume of Wigner functions $\protect\delta $
of the SPCCSs as a function of the catalysis parameter $\Lambda $ for
different input coherent state $\left\vert \protect\alpha \right\rangle $
with $\protect\alpha =1+i$ (red line), $\protect\alpha =2$ (blue line) and $%
\protect\alpha =2.7$ (green line).}
\end{figure}

On the other hand, the negative volume of the WF defined by 
\begin{equation}
\delta =\frac{1}{2}[\int_{-\infty }^{\infty }\int_{-\infty }^{\infty
}dqdp\left\vert W\left( q,p\right) \right\vert -1].  \label{4-3}
\end{equation}%
is a good indicator of non-classicality for a quantum state \cite{25}. In
Fig.8, we plot the negative volume $\delta $ of WF as\ a function of
different catalysis parameter $\Lambda $\ for several SPCCSs with $\alpha
=1+i $, $\alpha =2$, and $\alpha =2.7$. In addition, the negative areas are
modulated not only by the input parameter $\alpha $, but also by the
catalysis parameter $\Lambda $. It is found that for every given input $%
\left\vert \alpha \right\rangle $, there exist a maximum volume $\delta
_{\max }$ in a moderate catalysis parameter $\Lambda $. For instance, $%
\delta _{\max }$ is found at around $\Lambda =0.5$ for $\alpha =1+i$, at
around $\Lambda =0.25$ for $\alpha =2$, and at around $\Lambda =0.125$ for $%
\alpha =2.7$. Hence by adjusting the catalysis parameter $\Lambda $, the
optimal nonclassicality (i.e $\delta _{\max }$) can be obtained for a given
input coherent state $\left\vert \alpha \right\rangle $. In addition, we
also show that two extreme cases, i.e., $\delta _{\left\vert \alpha
\right\rangle }=0$ for $\Lambda =0$ and $\delta _{\left\vert 1\right\rangle
}=0.211081$ for $\Lambda =1$, are always right, as expected.

\section{Conclusion}

In summary, we have proved the equivalent effect between the lossless beam
splitter and the nondegenerate parametric amplifier in quantum state
engineering. We report theoretical preparation and nonclassical properties
of a kind of new non-Gaussian quantum states, i.e. single-photon catalyzed
coherent states (SPCCSs). These states are generated by operating
single-photon quantum-optical catalysis on a coherent state. Lossless beam
splitter and nondegenerate parametric amplifier are used as the catalyzed
devices respectively. We prove that the catalyzed coherent states are
actually the same quantum states. Although the success probabilities of the
detection are different, the effects of BS and PA are equivalent once the
detections are succeed. The quantum properties of the catalyzed states, such
as photon number distribution, quadrature squeezing effect, Mandel Q
parameter, autocorrelation function and Wigner function are investigated.
Simple extensions of the catalysis scheme allow for the preparation of more
sophisticated quantum states.

\begin{acknowledgments}
This work was supported by the National Nature Science Foundation of China
(Grants No. 11264018 and No. 11447002) and the Natural Science Foundation of
Jiangxi Province of China (Grants No. 20142BAB202001 and No. 20151BAB202013)
\end{acknowledgments}

\textbf{Appendix A: Character of the SPCCS generated by BS}

We give the detailed procedure of derivating the explicit form of the SPCCS $%
\left\vert \alpha _{c}\right\rangle _{BS}$ by BS. Noting the coheret state $%
\left\vert \alpha _{a}\right\rangle =e^{-\left\vert \alpha \right\vert
^{2}/2}e^{\alpha a^{\dag }}\left\vert 0_{a}\right\rangle $ and the Fock
state $\left\vert 1_{b}\right\rangle =\frac{d}{ds_{1}}e^{s_{1}b^{\dag
}}\left\vert 0_{b}\right\rangle |_{s_{1}=0}$, we rewrite $\left\vert \alpha
_{c}\right\rangle _{BS}$\ as 
\begin{eqnarray*}
\left\vert \alpha _{c}\right\rangle _{BS} &=&\frac{e^{-\left\vert \alpha
\right\vert ^{2}/2}}{\sqrt{p_{BS}}}\frac{d^{2}}{dt_{1}ds_{1}}e^{\left(
\alpha t+s_{1}r\right) a^{\dag }}\left\vert 0_{a}\right\rangle \\
&&\times \left\langle 0_{b}\right\vert e^{t_{1}b}e^{\left( s_{1}t-\alpha
r\right) b^{\dag }}\left\vert 0_{b}\right\rangle |_{s_{1}=t_{1}=0}, \\
&=&\frac{e^{-\left\vert \alpha \right\vert ^{2}/2}}{\sqrt{p_{BS}}}\frac{d^{2}%
}{dt_{1}ds_{1}}e^{\allowbreak ts_{1}t_{1}-r\alpha t_{1}+a^{\dag }t\alpha
+a^{\dag }rs_{1}}\left\vert 0_{a}\right\rangle |_{s_{1}=t_{1}=0} \\
&=&\frac{e^{-\left\vert \alpha \right\vert ^{2}/2}}{\sqrt{p_{BS}}}\left(
\allowbreak t-a^{\dag }r^{2}\alpha \right) \exp \left( \allowbreak t\alpha
a^{\dag }\right) \left\vert 0_{a}\right\rangle .
\end{eqnarray*}%
Thus the explicit form in Eq.(\ref{1-1}) is obtained. In addition, its
density operator can be read as%
\begin{align*}
\rho _{c-BS}& =\frac{e^{-\left\vert \alpha \right\vert ^{2}}}{p_{BS}}\frac{%
d^{4}}{dt_{1}ds_{1}dt_{2}ds_{2}}e^{ts_{1}t_{1}-r\alpha
t_{1}+ts_{2}t_{2}-r\alpha ^{\ast }t_{2}} \\
& \times e^{a^{\dag }t\alpha +a^{\dag }rs_{1}}\left\vert 0_{a}\right\rangle
\left\langle 0_{a}\right\vert e^{at\alpha ^{\ast
}+ars_{2}}|_{s_{1}=t_{1}=s_{2}=t_{2}=0},
\end{align*}%
and then its success probability $p_{BS}$ is obtained%
\begin{eqnarray*}
p_{BS} &=&e^{-\left\vert \alpha \right\vert ^{2}}\frac{d^{4}}{%
dt_{1}ds_{1}dt_{2}ds_{2}}e^{\allowbreak ts_{1}t_{1}-r\alpha
t_{1}+ts_{2}t_{2}-r\alpha ^{\ast }t_{2}} \\
&&\times e^{\allowbreak r^{2}s_{1}s_{2}+t^{2}\alpha \alpha ^{\ast }+rt\alpha
s_{2}+rts_{1}\allowbreak \alpha ^{\ast }}|_{s_{1}=t_{1}=s_{2}=t_{2}=0}.
\end{eqnarray*}

\textbf{Appendix B: Character of the SPCCS generated by PA}

We give the detailed procedure of derivating the explicit form of the SPCCS $%
\left\vert \alpha _{c}\right\rangle _{PA}$ by PA. Noting the integration of
squeezing operator $S\left( \lambda \right) =\int \frac{d^{2}\eta }{\mu \pi }%
\left\vert \frac{\eta }{\mu }\right\rangle \left\langle \eta \right\vert $
(with $\mu =e^{\lambda }$ and $\left\vert \eta \right\rangle =e^{-\frac{%
\left\vert \eta \right\vert ^{2}}{2}+\eta a^{\dag }-\eta ^{\ast }b^{\dag
}+a^{\dag }b^{\dag }}\left\vert 0_{a},0_{b}\right\rangle $) and the Fock
state $\left\vert 1_{b}\right\rangle =\frac{d}{ds_{1}}e^{s_{1}b^{\dag
}}\left\vert 0_{b}\right\rangle |_{s_{1}=0}$, we rewrite $\left\vert \alpha
_{c}\right\rangle _{PA}$\ as%
\begin{eqnarray*}
&&\left\vert \alpha _{c}\right\rangle _{PA} \\
&=&\frac{1}{\sqrt{p_{PA}}}\frac{d^{2}}{ds_{1}dt_{1}} \\
&&\times \int \frac{d^{2}\eta }{\mu \pi }\left\langle 0_{b}\right\vert
e^{t_{1}b}e^{-\frac{\left\vert \eta \right\vert ^{2}}{2\mu ^{2}}+\frac{\eta 
}{\mu }a^{\dag }-\frac{\eta ^{\ast }}{\mu }b^{\dag }+a^{\dag }b^{\dag
}}\left\vert 0_{a},0_{b}\right\rangle \\
&&\times \left\langle 0_{a},0_{b}\right\vert e^{-\frac{\left\vert \eta
\right\vert ^{2}}{2}+\eta ^{\ast }a-\eta b+ab}\left\vert \alpha
_{a}\right\rangle e^{s_{1}b^{\dag }}\left\vert 0_{b}\right\rangle
|_{s_{1}=t_{1}=0} \\
&=&\frac{\varkappa e^{-\left\vert \alpha \right\vert ^{2}/2}}{\sqrt{p_{PA}}}%
\frac{d^{2}}{ds_{1}dt_{1}}e^{-\kappa \alpha s_{1}+\varkappa
s_{1}t_{1}+\kappa t_{1}a^{\dag }+\varkappa \alpha a^{\dag }}\left\vert
0_{a}\right\rangle |_{s_{1}=t_{1}=0} \\
&=&\frac{\varkappa e^{-\left\vert \alpha \right\vert ^{2}/2}}{\sqrt{p_{PA}}}%
\left( \varkappa -\alpha a^{\dag }\kappa ^{2}\right) e^{\varkappa \alpha
a^{\dag }}\left\vert 0_{a}\right\rangle .
\end{eqnarray*}%
Thus the explicit form in Eq.(\ref{1-2}) is obtained. Meanwhile, its density
operator can be read as%
\begin{eqnarray*}
\rho _{c-PA} &=&\frac{\varkappa ^{2}e^{-\left\vert \alpha \right\vert ^{2}}}{%
p_{PA}}\frac{d^{4}}{ds_{1}dt_{1}ds_{2}dt_{2}} \\
&&\times e^{-\kappa \alpha s_{1}+\varkappa s_{1}t_{1}-\kappa \alpha ^{\ast
}s_{2}+\varkappa s_{2}t_{2}} \\
&&\times e^{\left( \kappa t_{1}+\varkappa \alpha \right) a^{\dag
}}\left\vert 0\right\rangle \left\langle 0\right\vert e^{\left( \kappa
t_{2}+\varkappa \alpha ^{\ast }\right) a}|_{s_{1}=t_{1}=s_{2}=t_{2}=0},
\end{eqnarray*}%
and then its success probability $p_{PA}$ is obtained%
\begin{eqnarray*}
p_{PA} &=&\varkappa ^{2}e^{-\kappa ^{2}\left\vert \alpha \right\vert ^{2}}%
\frac{d^{4}}{ds_{1}dt_{1}ds_{2}dt_{2}}e^{\varkappa \left(
s_{1}t_{1}+s_{2}t_{2}\right) +t_{1}t_{2}\allowbreak \kappa ^{2}} \\
&&\times e^{\left( \varkappa t_{2}-s_{1}\right) \alpha \kappa +\left(
\allowbreak \varkappa t_{1}-s_{2}\right) \alpha ^{\ast }\kappa
}|_{s_{1}=t_{1}=s_{2}=t_{2}=0}.
\end{eqnarray*}

\textbf{Appendix C: Statistical quantities of the SPCCSs}

In order to explore the statistical quantities of the SPCCS, we give the
general form of expectation value $\left\langle a^{\dag k}a^{l}\right\rangle 
$ as follows 
\begin{eqnarray*}
&&\left\langle a^{\dag k}a^{l}\right\rangle _{BS} \\
&=&\frac{e^{-r^{2}\left\vert \alpha \right\vert ^{2}}}{p_{BS}}\frac{d^{4+k+l}%
}{dt_{1}ds_{1}dt_{2}ds_{2}d\mu ^{k}d\nu ^{l}} \\
&&\times e^{\allowbreak ts_{2}t_{2}-r\alpha ^{\ast }t_{2}+\allowbreak
ts_{1}t_{1}-r\alpha t_{1}+\allowbreak t\alpha \nu +r\nu s_{1}+\allowbreak
r^{2}s_{1}s_{2}+r\mu s_{2}} \\
&&\times e^{+t\mu \alpha ^{\ast }+\allowbreak rt\alpha s_{2}+rts_{1}\alpha
^{\ast }}|_{s_{1}=t_{1}=s_{2}=t_{2}=\mu =\nu =0,}
\end{eqnarray*}%
and%
\begin{eqnarray*}
&&\left\langle a^{\dag k}a^{l}\right\rangle _{PA} \\
&=&\frac{\varkappa ^{2}e^{-\kappa ^{2}\left\vert \alpha \right\vert ^{2}}}{%
p_{PA}}\frac{d^{4+k+l}}{dt_{1}ds_{1}dt_{2}ds_{2}d\mu ^{k}d\nu ^{l}} \\
&&\times e^{\left( \nu \alpha +\mu \allowbreak \alpha ^{\ast }\right)
\varkappa +\left( s_{1}t_{1}+s_{2}t_{2}\right) \varkappa -\left( s_{1}\alpha
+\allowbreak s_{2}\alpha ^{\ast }\right) \kappa +\left( t_{2}\alpha
\allowbreak +t_{1}\alpha ^{\ast }\right) \varkappa \kappa } \\
&&\times e^{+\left( \mu t_{2}+\nu t_{1}\right) \kappa +t_{1}t_{2}\kappa
^{2}}|_{s_{1}=t_{1}=s_{2}=t_{2}=r_{1}=r_{2}=0}.
\end{eqnarray*}%
where we remain their differential forms. For different $k$ and $l$, we have%
\begin{eqnarray*}
\left\langle a^{\dag }\right\rangle _{\Lambda } &=&\left\langle
a\right\rangle _{\Lambda }^{\ast }=\frac{I_{1}\left( \alpha ,\Lambda \right) 
}{I_{0}\left( \alpha ,\Lambda \right) }\sqrt{\left( 1-\Lambda \right) }%
\alpha ^{\ast }, \\
\left\langle a^{\dag 2}\right\rangle _{\Lambda } &=&\left\langle
a^{2}\right\rangle _{\Lambda }^{\ast }=\frac{I_{2}\left( \alpha ,\Lambda
\right) }{I_{0}\left( \alpha ,\Lambda \right) }\left( 1-\Lambda \right)
\alpha ^{\ast 2}, \\
\left\langle a^{\dag }a\right\rangle _{\Lambda } &=&\frac{I_{3}\left( \alpha
,\Lambda \right) }{I_{0}\left( \alpha ,\Lambda \right) }\left\vert \alpha
\right\vert ^{2}, \\
\left\langle a^{\dag 2}a^{2}\right\rangle _{\Lambda } &=&\frac{I_{4}\left(
\alpha ,\Lambda \right) }{I_{0}\left( \alpha ,\Lambda \right) }\left(
1-\Lambda \right) \left\vert \alpha \right\vert ^{4},
\end{eqnarray*}%
where the functions are defined as 
\begin{eqnarray*}
I_{0}\left( \alpha ,\Lambda \right) &=&\left( 1-\Lambda \right) +\Lambda
\left( 3\Lambda -2\right) \allowbreak \left\vert \alpha \right\vert ^{2} \\
&&+\Lambda ^{2}\left( 1-\Lambda \right) \allowbreak \allowbreak \left\vert
\alpha \right\vert ^{4}, \\
I_{1}\left( \alpha ,\Lambda \right) &=&\left( 1-2\Lambda \right) +2\Lambda
\left( 2\Lambda -1\right) \left\vert \alpha \right\vert ^{2} \\
&&+\Lambda ^{2}\left( 1-\Lambda \right) \left\vert \alpha \right\vert ^{4},
\\
I_{2}\left( \alpha ,\Lambda \right) &=&\allowbreak \left( 1-3\Lambda \right)
+\Lambda \left( 5\Lambda -2\right) \left\vert \alpha \right\vert ^{2} \\
&&+\Lambda ^{2}\left( 1-\Lambda \right) \left\vert \alpha \right\vert ^{4},
\\
I_{3}\left( \alpha ,\Lambda \right) &=&\left( 2\Lambda -1\right)
^{2}+\Lambda \left( 1-\Lambda \right) \left( \allowbreak 5\Lambda -2\right)
\left\vert \alpha \right\vert ^{2} \\
&&+\Lambda ^{2}\left( 1-\Lambda \right) ^{2}\left\vert \alpha \right\vert
^{4}, \\
I_{4}\left( \alpha ,\Lambda \right) &=&\left( 3\Lambda -1\right)
^{2}+\Lambda \left( 1-\Lambda \right) \left( \allowbreak 7\Lambda -2\right)
\left\vert \alpha \right\vert ^{2} \\
&&+\Lambda ^{2}\left( 1-\Lambda \right) ^{2}\left\vert \alpha \right\vert
^{4}.
\end{eqnarray*}%
with $\Lambda =r^{2}$ for the BS catalysis and $\Lambda =\kappa ^{2}$ for
the PA catalysis.

\end{document}